\documentclass[letterpaper]{article} 
\usepackage{aaai2027}  
\usepackage[hyphens]{url}  
\usepackage{graphicx} 
\usepackage{natbib}  
\usepackage{caption} 
\usepackage{algorithm}
\usepackage{algorithmic}
\usepackage{amsmath}
\usepackage{amssymb}
\usepackage{multirow}

\usepackage{comment}

\usepackage{color}

\usepackage{xcolor}
\usepackage{makecell}
\definecolor{cup}{RGB}{197,48,48}    
\definecolor{cdown}{RGB}{26,122,72}  

\newcommand{\dn}[1]{\textcolor{black}{\scriptsize$-#1\%$}}
\usepackage{xspace}
\newcommand{\tool}{\textit{ECLoop}\xspace}

\newcommand{\peter}[1]{\textcolor{blue}{{\it [peter says: #1]}}}

\newcommand{\ys}[1]{\textcolor{orange}{{\it [yisen says: #1]}}}

\definecolor{cgreen}{HTML}{228B22}
\usepackage{newfloat}
\usepackage{listings}
\DeclareCaptionStyle{ruled}{labelfont=normalfont,labelsep=colon,strut=off} 
\floatstyle{ruled}
\newfloat{listing}{tb}{lst}{}
\floatname{listing}{Listing}

\usepackage{booktabs}

\nocopyright
\title{Preventing Premature Commitment in Coding Agents with an Evidence-Conditioned Execution Layer}

\author{
    Yisen Xu\textsuperscript{\rm 1,\rm 2}\equalcontrib,
    Chenglin Li\textsuperscript{\rm 1}\equalcontrib,
    Zehao Wang\textsuperscript{\rm 1},
    Jinqiu Yang\textsuperscript{\rm 2},
    Tse-Hsun (Peter) Chen\textsuperscript{\rm 1}\corresponding
}
\affiliations{
    \textsuperscript{\rm 1}SPEAR Lab, Concordia University, Montreal, Quebec, Canada\\
    \textsuperscript{\rm 2}O-RISA Lab, Concordia University, Montreal, Quebec, Canada\\
    yisen.xu@mail.concordia.ca, chenglin.li@mail.concordia.ca, w\_zeha@encs.concordia.ca,\\
    jinqiu.yang@concordia.ca, peterc@encs.concordia.ca
}

\begin{document}

\maketitle

\begin{abstract}
LLM-based coding agents often edit source code or submit patches
before examining enough repository evidence to justify the change,
a failure pattern we call \emph{premature commitment}.
We present \tool, an execution layer that interposes between
the agent and the repository to enforce evidence-conditioned execution.
For each task, \tool uses the issue description and repository structure
to compile a set of conditions specifying what the agent should observe
before each type of code modification or patch submission.
During execution, \tool tracks which conditions the agent's runtime
trajectory has satisfied and postpones any proposed action whose
required conditions remain unmet.
Evaluated on all 500 instances of SWE-bench Verified with two
language models and two agent scaffolds, \tool raises Pass@1
by 4.8--11.8 percentage points without model retraining or
scaffold changes.
Ablation experiments show that each of \tool's three operations
contributes distinct value and that structured evidence conditions
outperform an equivalent natural-language summary.
These gains come at no additional inference cost: by redirecting
the agent before it pursues unsupported actions, \tool lowers
average token consumption by up to 12.1\%.

\end{abstract}


\section{Introduction}
Coding agents may edit code or submit a patch before examining enough repository evidence to support that action. An edit may appear plausible after inspecting the target function, even though an unseen caller, related implementation, test, or behavioral constraint would change that judgment. Likewise, a patch may appear ready for submission after one test passes while other relevant behavior remains unexamined. Thus, beyond generating a plausible repair~\cite{autocoderover,agentless}, an agent must determine whether it has observed enough evidence for the proposed action to proceed. We call executing an edit or submission before the relevant evidence has been examined a \emph{premature commitment}.

Existing coding agents provide no mechanism to detect or prevent premature commitment. At each step, the agent decides what action to take based on the evidence gathered so far, but it cannot assess whether that evidence is sufficient to justify a commitment action such as editing a file or submitting a patch~\cite{react,yang2024sweagent,codex,openhands,repairagent}. Prompt-level instructions can encourage the agent to investigate more broadly before editing, but they cannot verify that the relevant evidence has actually been gathered or prevent an unsupported action from executing~\cite{fixedbench,agentc,agentspec}. As a result, whether a commitment action proceeds depends entirely on the agent's own judgment, with no independent check against the evidence collected so far.

\begin{figure*}
    \centering
    \includegraphics[width=\textwidth]{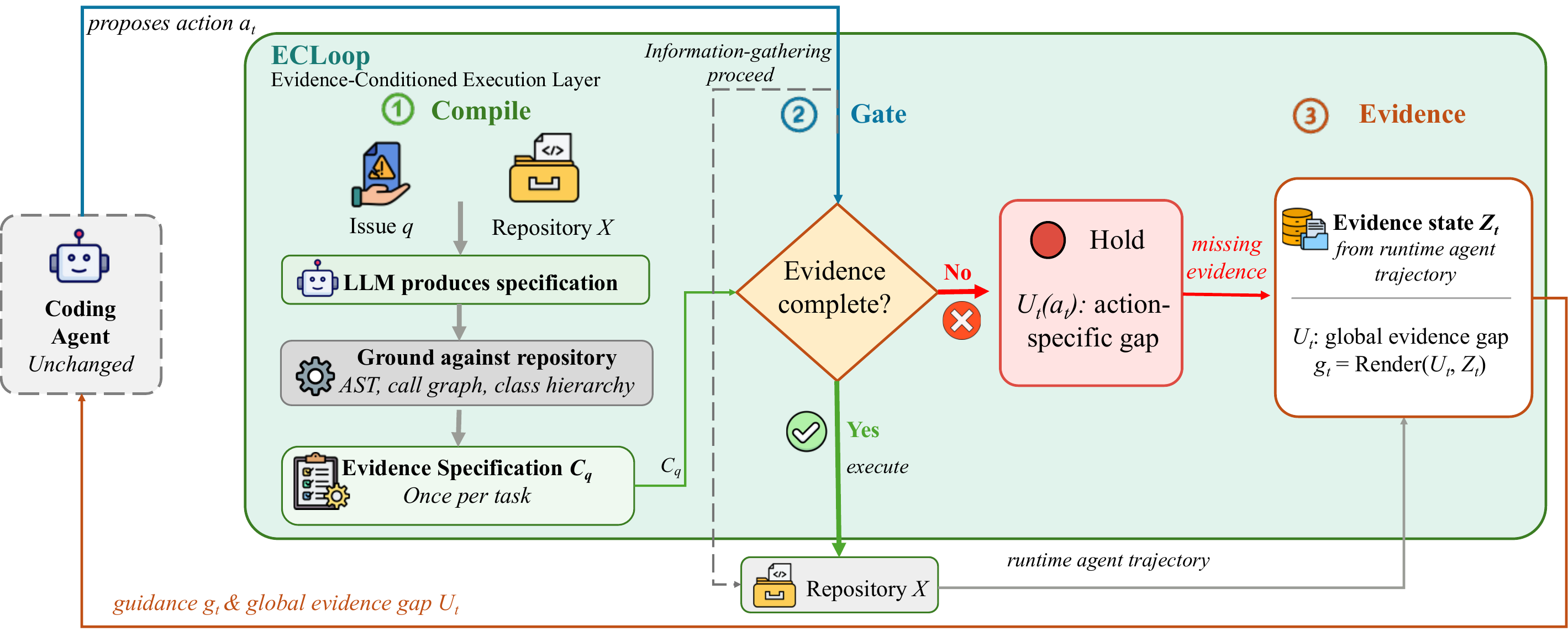}
   \caption{\textbf{\tool\ overview.}
    \tool\ interposes between the coding agent (left) and the
    repository: (1)~compiling evidence conditions from the issue once
    per task, (2)~gating commitment actions whose action-specific
    evidence gap remains nonempty, and (3)~returning the global
    evidence gap as guidance until all conditions are met.
    Information-gathering actions proceed freely.}
    \label{fig:ecloop-overview}
\end{figure*}

We present \tool, an execution layer that interposes between the agent's proposed actions and their execution (Figure~\ref{fig:ecloop-overview}). Before each commitment action, \tool checks whether the agent has gathered sufficient evidence to justify the proposed step. If relevant evidence is missing, the action is held, and the unsatisfied conditions are returned to the agent to guide further investigation. \tool controls only whether a proposed commitment is sufficiently supported to proceed. The agent remains responsible for how it investigates the repository and constructs the repair. As an execution layer added to an existing agent, \tool requires neither model retraining nor changes to the agent's action-selection process.

Given an issue and repository, \tool first uses a language model to compile an evidence specification from the issue description and repository structure. The specification contains conditions describing the observable evidence that should be collected before particular commitment actions proceed, such as inspecting relevant callers before editing a function or checking the available tests before submitting a patch. 
As the agent investigates the repository, \tool parses each command and tool output in the runtime agent trajectory into structured events, and applies deterministic satisfaction checks to check which conditions these events satisfy, without invoking a language model. \tool maintains a \emph{global evidence gap}, consisting of all currently unsatisfied conditions, and adds this gap to the agent's context to guide further investigation.


When the agent proposes a commitment action, \tool derives an \emph{action-specific evidence gap} containing only the unsatisfied conditions applicable to that action. The action is postponed if this gap is nonempty and allowed to proceed otherwise. The global gap therefore guides investigation, while the action-specific gap controls execution. \tool guarantees that a commitment proceeds only after the required observable evidence has been recorded, although the agent may still misinterpret that evidence.


We evaluate \tool on all 500 instances of SWE-bench Verified ~\cite{jimenez2024swebench} across two language models, \textbf{MiniMax-M2.5}~\cite{Minimax-m2.5} and \textbf{GPT-5-mini}~\cite{gpt-5-mini}, and two agent scaffolds, \textbf{mini-swe-agent v2}~\cite{yang2024sweagent} and OpenAI's \textbf{Codex CLI}~\cite{codex}. Across all four configurations, \tool improves Pass@1 by 4.8--11.8 percentage points. With mini-swe-agent v2, it raises Pass@1 from 75.8\% to 80.6\% under MiniMax-M2.5 and from 56.2\% to 68.0\% under GPT-5-mini. Integrating \tool with Codex CLI produces comparable gains, improving MiniMax-M2.5 from 74.8\% to 79.8\% and GPT-5-mini from 40.4\% to 50.8\%. 

Notably, these accuracy improvements do not come at higher inference cost: by holding unsupported actions before they trigger unproductive trajectories, \tool reduces average token usage by up to 12.1\%.

In summary, this work makes the following contributions:
\begin{itemize}
\item We identify \emph{premature commitment} as a distinct failure mode in coding agents and formulate evidence-conditioned execution, which separates proposing an action from deciding whether sufficient evidence has been collected for it to proceed.
\item We introduce \tool, an execution control layer that tracks task-specific evidence, guides the agent using what remains missing, and checks relevant evidence before edits and final submissions.

\item On all 500 SWE-bench Verified instances, \tool improves Pass@1 for both GPT-5-mini and MiniMax-M2.5 across mini-swe-agent v2 and Codex CLI, with gains of 4.8--11.8 percentage points and a best Pass@1 of 80.6\%. 
Across all four model--scaffold configurations, \tool also reduces token usage by 1.4--12.1\% and inference cost by 1.5--10.2\%.
\end{itemize}

\section{Related Work}
\subsection{Evidence Gathering in Coding Agents}
\label{sec:rw-evidence}

Recent studies show that coding agents frequently produce patches without sufficient repository evidence. Agents generate edits for already-fixed issues~\cite{fixedbench}, miss relevant repository context that would change the repair~\cite{swe_explore}, and produce incomplete or test-overfitted patches due to insufficient issue and code understanding~\cite{refine}. These findings show that repair quality depends not only on patch generation but also on what evidence the agent gathers before committing to a change.

Several systems address this by building evidence checks into multi-stage repair pipelines. \textsc{EviACT}~\cite{eviact} coordinates a retrieval scaffold for localization, a compile gate for rejecting invalid patches, and a test-driven gate for validating target-test recovery across a staged pipeline. \textsc{SWE-Doctor}~\cite{swedoctor} requires a reproduction test to pass before patch generation, following work that treats reproduction tests as executable specifications~\cite{swtbench,otter,eotterjava,reproagent}. In these systems, the required evidence is tied to specific tools and artifacts, such as
compiler outcomes, reproduction tests, or target-test results, and is
checked at predefined points in the workflow. 

\tool instead derives evidence conditions from each issue, so different tasks require different evidence rather than passing through the same fixed gates. It evaluates these conditions at every commitment boundary and operates as an execution layer on top of an existing scaffold, without modifying the scaffold's action-selection policy.

\subsection{Runtime Enforcement for LLM Agents}
\label{sec:rw-enforcement}

Runtime enforcement systems constrain agent behavior during
execution. \textsc{AgentSpec} enforces specified constraints on individual actions, while \textsc{Agent-C} enforces temporal constraints over sequences of tool calls~\cite{agentspec,agentc}. \textsc{ProbGuard} predicts trajectories that may reach unsafe states~\cite{probguard}. Other systems generate, adapt, or verify guard policies~\cite{veriguard,agrail,guardagent}, while \textsc{MI9} and \textsc{NeMo Guardrails} apply controls across multiple system layers~\cite{mi9,nemoguardrails}. \textsc{TRIAD} extends proceed-or-refuse enforcement with an update decision and verbal feedback that helps the agent revise unsafe plans~\cite{triad}.

Despite their different implementations, these systems primarily enforce \emph{behavioral compliance}: whether an action or trajectory satisfies a safety, security, or governance policy. However, an edit may violate no such policy and follow every required tool order while still being premature because the agent has not established that the change is needed.

\tool addresses this distinct failure mode by enforcing \emph{evidence-conditioned execution}. It checks whether the agent's trajectory contains the observations needed to justify the current repair task, rather than whether the action satisfies a fixed policy. Although \textsc{TRIAD} also provides verbal feedback, its feedback remediates unsafe plans, whereas \tool's feedback directs the agent toward the specific repository investigation still needed.

\subsection{Intervening on Agent Trajectories}
\label{sec:rw-intervention}

Several methods improve agent trajectories by helping agents choose better actions. \textsc{ReAct}~\cite{react} interleaves reasoning traces with actions, allowing the agent to update its plan based on new observations. \textsc{Reflexion}~\cite{reflexion} uses verbal self-reflection to improve later trials. Process reward models score intermediate steps to guide better action choices~\cite{agentprm,swetrace}. These methods primarily improve which action the agent proposes, yet agents may recognize that a tool is needed and still fail to call it during generation~\cite{knowwhentocall}.

Premature commitment is related to the classical least-commitment principle, which delays decisions until available constraints require them~\cite{leastcommitment}. Direct intervention also carries risk: it may recover failing trajectories but disrupt ones that would otherwise succeed~\cite{interventionparadox}.

\tool instead addresses a different question: whether the agent is ready to commit. It does not choose the agent's next action, but checks whether the agent has gathered the task-specific evidence needed to justify an edit or final submission.

\section{\tool: Evidence-Conditioned Execution}
\label{sec:method}


A coding agent may propose a plausible action before completing the investigation needed to support it (i.e., the \textit{evidence}). 
For example, it may identify a likely edit while relevant callers, related implementations, or failure conditions remain unexamined. Conventional agent runtimes
typically execute a proposed action
without separately assessing whether the available evidence is sufficient. This conflates two decisions: whether the action is a plausible next step and whether the agent is ready to commit to it. 

We introduce \tool, an evidence-conditioned execution layer that separates action proposal from execution readiness. The coding agent continues to decide what to do next, while \tool tracks task-specific evidence from the runtime agent trajectory and determines whether actions that modify or finalize the solution
should proceed. \textit{\textbf{Information-gathering actions}}, including file reads, code searches, and diagnostic commands, proceed without restriction. In contrast, \emph{\textbf{commitment actions}}, such as source-code edits and final submission, proceed only when supported by the relevant evidence.

\tool maintains two views of the evidence still missing from the
ongoing investigation, which we refer to as the \emph{evidence gap}. The \emph{global evidence gap} is added to the model context so the agent can see what evidence the task still needs. When the agent proposes a commitment action, \tool derives an \emph{action-specific evidence gap} containing only the unsatisfied conditions relevant to that action, and gates the action based on whether those conditions have been met. Beyond these two interventions, the underlying agent remains unchanged.

As illustrated in Figure~\ref{fig:ecloop-overview}, \tool operates in three stages. It first analyzes the issue and repository to compile task-specific conditions describing the evidence needed before a commitment action can proceed. During execution, it uses the runtime agent trajectory to determine which conditions have been satisfied and adds the remaining unsatisfied evidence gap to the model context. When the agent proposes an edit or final submission, \tool checks only the conditions relevant to that action. If any remain unsatisfied, it holds the action and returns unsatisfied conditions to guide the agent’s next action.

\subsection{Compiling the Evidence Specification}
\label{sec:evidence-compilation}

Given an issue $q$ and repository $X$, \tool constructs a task-specific evidence specification that defines what the agent must establish before a commitment action can proceed. This compilation separates two questions: which evidence is needed for the task, and how its satisfaction can be determined from the repository and the runtime agent trajectory. We present \tool in the context of software repair; applying it to other domains would require domain-specific representations of evidence and corresponding procedures for evaluating them.

To identify the evidence relevant to the task, \tool uses a language model to analyze the issue together with structural information extracted from the repository and produce a task-specific evidence specification:
\begin{equation}
C_q=\{c_1,\ldots,c_m\},
\label{eq:evidence-specification}
\end{equation}
where each condition $c_i$ specifies an investigation requirement and the commitment actions to which it applies. At this stage, a condition may specify an abstract program role without yet resolving it to a concrete repository entity. For example, ``the \textit{function} to be edited,'' ``\textit{its} callers,'' and ``\textit{related} implementations'' describe what must be inspected, but not which functions or files satisfy those roles in the current task. The specification therefore defines what evidence is needed without assuming that the relevant entities are already known. 

\paragraph{Formal evidence representation.}
Free-form instructions are inefficient for execution control because they do not explicitly specify how required evidence should be tracked, when it applies, or how its completion should be determined.  \tool\ therefore represents each evidence condition as a structured specification~\cite{agentspec,agentsama}, explicitly recording the relevant program entity, the observable evidence required, the commitment action to which it applies, and a satisfaction predicate. 

\tool\ converts each abstract condition in $C_q$ into a repository-specific condition by identifying the concrete program entities to which it applies. To perform this resolution, \tool uses abstract syntax tree traversal to locate relevant functions and classes, the call graph to identify caller--callee relationships, and the class hierarchy to identify related or overriding implementations. Each condition~$c_i$ is then represented as
\begin{equation}
\phi_i
=
\langle b_i,v_i,R_i,\operatorname{sat}_i\rangle,
\label{eq:evidence-condition}
\end{equation}
where $b_i$ specifies the type of commitment action at which the condition is checked, $v_i$ denotes the concrete program entity to which it applies, $R_i$ specifies the event pattern that must appear in the runtime agent trajectory, and $\operatorname{sat}_i(Z_t)\in\{0,1\}$ indicates whether the event pattern $R_i$ has been matched in the current evidence state~$Z_t$.

This representation allows each condition to be evaluated against the runtime agent trajectory, presented to the agent as guidance, and checked when a commitment action is proposed.  By preserving individual conditions throughout execution, \tool\ avoids the ambiguity introduced by compressing the remaining evidence into an unconstrained natural-language summary.

Satisfaction is computed from the runtime agent trajectory, such as inspected
code locations, executed commands, and diagnostic outcomes, rather than
from the model claiming that it completed an investigation. Grounding
succeeds only when the references in a condition can be resolved and its
satisfaction can be checked:
\begin{equation}
    \operatorname{Ground}(c_i,X)
    =
    \begin{cases}
        \phi_i,
        & \text{if $c_i$ resolves to checkable evidence},\\
        \bot,
        & \text{otherwise}.
    \end{cases}
    \label{eq:condition-grounding}
\end{equation}
Conditions that cannot be resolved to concrete repository entities are removed from the specification. The language model therefore proposes candidate conditions, but does not determine their execution-time enforcement. A condition becomes active only after its references have been resolved, and its satisfaction is evaluated deterministically from the runtime agent trajectory. Errors in the generated specification may reduce coverage, but they cannot introduce checks that the system cannot evaluate.

Some conditions can be resolved before execution because the issue already identifies the relevant program entity or failure. Others can be resolved only after the agent proposes a specific action. For example, a condition requiring caller inspection cannot identify the relevant callers until the agent proposes editing a function $f$. \tool then uses $f$ to retrieve its callers from the call graph and checks the runtime agent trajectory to determine whether they have been inspected.

Let $C_t$ denote the grounded conditions available by step $t$. Before
evaluating a proposed commitment action, \tool grounds any additional
conditions whose references can now be resolved from that action and
adds them to $C_t$. Thus, $C_q$ captures the task-specific
investigation requirements, while $C_t$ contains the concrete
conditions that can be evaluated during execution.

\subsection{Maintaining the Global Evidence Gap}
\label{sec:evidence-guidance}

At step $t$, \tool derives the current evidence state from the runtime agent trajectory:
\begin{equation}
    Z_t
    =
    \operatorname{Observe}(\tau_t).
    \label{eq:evidence-state}
\end{equation}
\tool represents each executed command and its output as observable events describing what the agent inspected, searched, modified, or executed. These events record the relevant code locations, symbols, files, line ranges, and execution outcomes. Each predicate $\operatorname{sat}_i$ specifies the event pattern required to satisfy its condition. For example, an inspection condition requires an event showing that the relevant code location was viewed, while a reproduction condition requires a recognized failure outcome before the edit. Hence, satisfaction is determined directly from the recorded trajectory using fixed, condition-specific criteria.

\tool evaluates the grounded conditions in $C_t$ against $Z_t$ and collects those that remain unsatisfied:
\begin{equation}
    U_t
    =
    \{
        \phi_i\in C_t
        \mid
        \operatorname{sat}_i(Z_t)=0
    \}.
    \label{eq:global-evidence-gap}
\end{equation}
We refer to $U_t$ as the \textit{\textbf{global evidence gap}}. It summarizes the task-relevant evidence that remains missing at this point in the trajectory.

\tool converts the remaining evidence gap into guidance for the agent:
\begin{equation}
g_t
=
\operatorname{Render}(U_t,Z_t),
\label{eq}
\end{equation}
where $g_t$ describes what still needs to be established and refers to concrete program entities when available. Conditions that have already been satisfied are omitted, so the agent receives only the evidence that remains missing rather than the full initial specification.

The model generates its next response conditioned on the issue, the runtime agent trajectory, and the current evidence gap:
\begin{equation}
y_{t,k}
\sim
p_\theta
\bigl(
\cdot
\mid
q,\tau_t,g_t,y_{t,<k}
\bigr).
\label{eq:evidence-conditioned-generation}
\end{equation}
Here, $\tau_t$ provides the execution history, while $g_t$ explicitly identifies the evidence that remains missing. Conditioning generation on $g_t$ directs the agent toward unresolved investigation requirements without prescribing a specific action or order. The agent therefore retains control over how to obtain the required evidence.

After an executed action $a_t$ returns observation $o_t$, \tool
updates the trajectory and evidence state:
\begin{equation}
    \tau_{t+1}
    =
    \tau_t\oplus\langle a_t,o_t\rangle,
    \qquad
    Z_{t+1}
    =
    \operatorname{Observe}(\tau_{t+1}).
    \label{eq:evidence-update}
\end{equation}
The updated state produces a new global evidence gap for the next
step. Evidence is therefore maintained as a dynamic execution state
rather than as a static instruction supplied only at the beginning of
the task.

\subsection{Gating Commitment by the Action-Specific Gap}
\label{sec:evidence-gated-execution}



The global evidence gap guides the agent toward what remains to be established, but guidance alone cannot prevent a premature commitment. When the model proposes a commitment action, \tool first uses that action to resolve any condition references that could not be identified earlier and adds the newly resolved conditions to $C_t$. It then checks whether the evidence required for that action is complete.

For a proposed action $a_t$, \tool derives the action-specific evidence gap:
\begin{equation}
    U_t(a_t)
    =
    \left\{
        \phi_i\in C_t
        \;\middle|\;
        \operatorname{applies}(\phi_i,a_t)
        \land
        \operatorname{sat}_i(Z_t)=0
    \right\},
    \label{eq:action-specific-gap}
\end{equation}
where $\operatorname{applies}(\phi_i,a_t)$ holds when $a_t$ matches the
action type $b_i$ and affects the entity $v_i$. Unlike the global gap
$U_t$, which guides the overall investigation, $U_t(a_t)$ contains only
the missing evidence relevant to the proposed action.


A commitment action may proceed only when its action-specific evidence gap is empty:
\begin{equation}
U_t(a_t)=\varnothing.
\label{eq:execution-gate}
\end{equation}
Non-commitment actions are unaffected by this check. When a commitment action has unmet evidence requirements, \tool postpones it and adds the missing evidence to the next model context. The agent may then continue the investigation, revise the proposed action, or pursue a different approach. Thus, \tool constrains when commitment is allowed without prescribing the subsequent action.

\paragraph{Execution control and guarantees.}
The global evidence gap guides the agent's investigation, new observations update the evidence state, and the action-specific gap determines whether a proposed commitment may proceed. As an execution control layer, \tool determines whether a proposed commitment action is ready to proceed without deciding which action the agent should propose. 

\tool enforces evidence completion within a bounded hold budget. When a commitment action's applicable conditions remain unsatisfied, \tool postpones the action and returns the missing evidence to the agent, repeating this up to three times for the same commitment target. If the budget is exhausted without the conditions being met, the action proceeds through an audited fallback in which any unsatisfied conditions remain recorded, preserving the distinction between an evidence-supported commitment and one allowed by the fallback.

\section{Experiments}
\label{sec:experiments}

We evaluate whether \tool improves repository-level issue resolution across large language models and agent scaffolds, how much execution overhead and token cost it adds, and how much each component contributes.

\paragraph{Benchmark and metric.}
We evaluate \tool on all 500 instances of \textbf{SWE-bench
Verified}~\cite{jimenez2024swebench}, a human-validated subset of
SWE-bench in which each instance pairs a real GitHub issue with a
repository snapshot and held-out tests.
We report \textbf{Pass@1}, the fraction of instances resolved by a
single agent run.
Because stronger models/agents leave fewer failures available to recover,
we also report the \textbf{relative reduction in unresolved instances}
(RRU)~\cite{RRU},
\[
\mathrm{RRU}
=
\frac{\mathrm{Pass@1}_{\text{\tool}}-
      \mathrm{Pass@1}_{\mathrm{base}}}
     {1-\mathrm{Pass@1}_{\mathrm{base}}}.
\]
For each comparison, statistical significance is assessed using an
exact McNemar test~\cite{McNemar1947} on the 500 paired binary outcomes.

\paragraph{Models.}
We use \textbf{MiniMax-M2.5} with high reasoning effort~\cite{Minimax-m2.5} and \textbf{GPT-5-mini}~\cite{gpt-5-mini}. Both offer strong coding performance at a cost that enables full-benchmark evaluation. We apply the same model configuration across all experiments. 

\paragraph{Agent scaffolds.}
Both models run on \textit{\textbf{mini-swe-agent v2}}~\cite{yang2024sweagent}, a lightweight scaffold released by the SWE-agent project~\cite{yang2024sweagent}. To test whether \tool also improves a full-featured production coding agent in addition to a minimal research scaffold, we additionally integrate \tool with \textit{\textbf{OpenAI Codex CLI}} v0.144.4~\cite{codex} and run both models.

\paragraph{Baselines and controlled comparison.}
\tool acts as an execution layer on top of the unchanged agent. Within each configuration, the baseline and \tool share the same model, prompt, tools, and scaffold. The only addition is that \tool compiles task-specific evidence conditions with the same model, exposes the global evidence gap~$U_t$, and checks the action-specific evidence gap~$U_t(a_t)$ before each commitment action proceeds.

In addition to the baseline agents, we compare \tool against Self-Refine~\cite{selfrefine}, an iterative refinement method in which the model reviews the agent's own patch and revises it up to three iterations. Self-Refine uses comparable additional inference but operates after the agent has already produced a patch, rather than gating commitment actions during the agent loop.

\begin{table}[t]
\centering
\setlength{\tabcolsep}{4pt}
\begin{tabular}{@{}ll rrr@{}}
\toprule
Model & Method & Pass@1 & $\Delta$ & RRU \\
\midrule
\multirow{3}{*}{GPT-5-mini}
  & Baseline
  & 56.2\% & --- & --- \\
  & Self-Refine
  & 54.8\%
  & $-1.4$
  & $-3.2\%$ \\
  & \tool
  & \textbf{68.0\%}
  & \textbf{+11.8}
  & \textbf{26.9\%} \\
\midrule
\multirow{3}{*}{MiniMax-M2.5\textsuperscript{\dag}}
  & Baseline
  & 75.8\% & --- & --- \\
  & Self-Refine
  & 74.0\%
  & $-1.8$
  & $-7.4\%$ \\
  & \tool
  & \textbf{80.6\%}
  & \textbf{+4.8}
  & \textbf{19.8\%} \\
\bottomrule
\end{tabular}

\caption{Main results on SWE-bench Verified (500 instances) with
mini-swe-agent~v2. Both \tool improvements are significant under an
exact McNemar test~\cite{McNemar1947} ($p<0.001$).}
\label{tab:main-results}

\smallskip
{\raggedright\footnotesize
\textsuperscript{\dag}High reasoning effort.
RRU = relative reduction in unresolved instances.
Self-Refine~\cite{selfrefine} uses up to three revision iterations.\par}
\end{table}

\begin{table}[t]
\centering
\setlength{\tabcolsep}{5pt}
\begin{tabular}{@{}l rrrr@{}}
\toprule
Model & Base & \tool & $\Delta$ & RRU \\
\midrule
GPT-5-mini
  & 40.4\%
  & \textbf{50.8\%}
  & \textbf{+10.4}
  & \textbf{17.4\%} \\
MiniMax-M2.5\textsuperscript{\dag}
  & 74.8\%
  & \textbf{79.8\%}
  & \textbf{+5.0}
  & \textbf{19.8\%} \\
\bottomrule
\end{tabular}

\caption{Results of integrating \tool into Codex~CLI on SWE-bench
Verified (500 instances). Both improvements are statistically
significant (two-sided exact McNemar test~\cite{McNemar1947}, $p<0.001$).}
\label{tab:transfer-results}

\smallskip
{\raggedright\footnotesize
\textsuperscript{\dag}High reasoning effort.
RRU = relative reduction in unresolved instances.\par}
\end{table}

\begin{table*}[t]
\centering

\setlength{\tabcolsep}{18pt}
\begin{tabular}{@{}l ll ll ll@{}}
\toprule
& \multicolumn{2}{c}{Avg.\ Tokens (K)}
& \multicolumn{2}{c}{Avg.\ Cost}
& \multicolumn{2}{c@{}}{Total Cost} \\
\cmidrule(lr){2-3} \cmidrule(lr){4-5} \cmidrule(l){6-7}
Model & Base & \tool & Base & \tool & Base & \tool \\
\midrule
\multicolumn{7}{@{}l}{\textit{mini-swe-agent v2}} \\[2pt]
\quad GPT-5-mini
  & 179 & 157\dn{12.1}
  & \$0.047 & \$0.042\dn{10.2}
  & \$23.60 & \$21.18 \\
\quad MiniMax-M2.5\textsuperscript{\dag}
  & 219 & 206\dn{6.2}
  & \$0.073 & \$0.069\dn{6.1}
  & \$36.64 & \$34.39 \\[4pt]
\multicolumn{7}{@{}l}{\textit{Codex~CLI}} \\[2pt]
\quad GPT-5-mini
  & 111 & 105\dn{5.4}
  & \$0.030 & \$0.029\dn{5.3}
  & \$15.16 & \$14.36 \\
\quad MiniMax-M2.5\textsuperscript{\dag}
  & 165 & 162\dn{1.4}
  & \$0.055 & \$0.054\dn{1.5}
  & \$27.54 & \$27.14 \\
\bottomrule
\end{tabular}
\caption{Execution efficiency on SWE-bench Verified (per-instance
averages over 500 instances; costs in USD).}
\label{tab:efficiency}
\smallskip
{\raggedright\footnotesize
\textsuperscript{\dag}High reasoning effort.\par}
\end{table*}

\subsection{\tool Improves Repair Across LLMs}

\tool improves both models under mini-swe-agent~v2 (Table~\ref{tab:main-results}). For GPT-5-mini, Pass@1 increases from 56.2\% to 68.0\% (+11.8 percentage points, pp), corresponding to an RRU of 26.9\%. For MiniMax-M2.5, Pass@1 increases from 75.8\% to 80.6\% (+4.8~pp), corresponding to an RRU of 19.8\%. Despite their different baseline performance, the two models achieve RRUs of 26.9\% and 19.8\%, showing that \tool provides substantial relative gains for lower- and higher-performing baselines. 

Compared with the baseline agents, \tool newly resolves 33 and 68 failures for MiniMax-M2.5 and GPT-5-mini, respectively, while regressing on only 9 successes in each case. The few regressions occur when the agent repeatedly proposes a commitment action that the gate holds for insufficient evidence, exhausting the hold limit of three; the fallback release then permits an under-supported action that produces an incorrect patch.

Integrating Self-Refine with the mini-swe-agent v2 slightly degrades Pass@1 in both models (by $-1.4$ and $-1.8$~pp), confirming that post hoc self-review cannot recover from decisions made on insufficient evidence. \tool avoids this by gating commitment actions before they execute, preventing premature commitment rather than attempting to correct it afterward.

\subsection{Integrating \tool into Codex CLI}
\label{sec:scaffold-transfer}

To evaluate whether \tool transfers beyond mini-swe-agent~v2, we integrate it into Codex CLI~\cite{codex} through its hook mechanism. The hook intercepts proposed commitment actions before execution, checks them against the accumulated evidence, and either permits the action or returns feedback identifying the missing observation. This integration requires no modification to Codex CLI's model, tool set, or action-selection policy.

\tool improves both models under Codex CLI (Table~\ref{tab:transfer-results}). For GPT-5-mini, Pass@1 increases by 10.4~pp, corresponding to an RRU of 17.4\%; \tool newly resolves 68 baseline failures while regressing on 16 successes. For MiniMax-M2.5, Pass@1 increases by 5.0~pp, corresponding to an RRU of 19.8\%; \tool newly resolves 35 failures while regressing on 10 successes. These Pass@1 gains closely match those under mini-swe-agent~v2 (+11.8 and +4.8~pp), showing that \tool provides similar improvements across different coding-agent scaffolds.

\subsection{Execution Efficiency and Cost}

Table~\ref{tab:efficiency} reports per-instance token usage and monetary cost averaged over all 500 instances. \tool\ reduces both token usages and cost in all four configurations, with cost savings of 1.5--10.2\% and token reductions of 1.4--12.1\%. The reported totals include the per-task specification-compilation call. In short, \tool achieves the Pass@1 gains without increasing inference cost. Although \tool introduces additional evidence checks, it reduces overall token use by redirecting the agent before unsupported commitment actions lead to longer, less productive trajectories. 

The largest savings occur in the weak baseline configuration. With mini-swe-agent v2 and GPT-5-mini, which has the baseline Pass@1 at 56.2\%, \tool reduces token use by 12.1\% and cost by 10.2\%. In contrast, with Codex~CLI and MiniMax-M2.5, which have the higher baseline Pass@1 at 74.8\%, the reductions are 1.4\% and 1.5\%, respectively. This pattern suggests that \tool may provide better efficiency gains when the baseline is more likely to spend computational resources on unsuccessful trajectories.

\subsection{Ablation Analysis}

\tool\ has three operations: \emph{guidance}, which communicates the remaining evidence gap to the agent; the \emph{commitment check}, which blocks a commitment action while its relevant evidence remains incomplete; and the \emph{evidence-state update}, which recomputes the evidence state after each observation. We ablate each on a fixed, randomly sampled subset of 100 instances with GPT-5-mini.

As shown in Table~\ref{tab:mechanism-ablation}, removing any single operation from full \tool\ (68 solved) degrades performance. The commitment check contributes the most ($-$10 pp to ~58\%), while the evidence-state update is nearly as important ($-$9 pp to ~59\%). Without evidence-state update, newly observed evidence is not reflected in the gaps, weakening both guidance and checking. Guidance adds a further 5 pp. All three together outperform every partial variant, confirming that they provide complementary benefits. Replacing the structured specification with a natural-language summary also drops the Pass@1 by 10 pp, to 58\%, falling below even no guidance (Pass@1 of 63\%).


\begin{table}[t]
\centering

\begin{tabular}{@{}l rr@{}}
\toprule
Configuration & Solved & $\Delta$ \\
\midrule
Full \tool                          & \textbf{68} & -- \\
\quad$-$\,guidance                  & 63 & \textcolor{black}{$-$5\phantom{0}} \\
\quad$-$\,commitment check          & 58 & \textcolor{black}{$-$10} \\
\quad$-$\,state update              & 59 & \textcolor{black}{$-$9\phantom{0}} \\
Baseline ($-$\,all)                 & 47 & \textcolor{black}{$-$21} \\
\midrule
\quad spec $\to$ natural language   & 58 & \textcolor{black}{$-$10} \\
\bottomrule
\end{tabular}
\caption{Ablation on a fixed 100-instance subset using mini-swe-agent v2 and GPT-5-mini.
Each row removes or replaces one component of full \tool.}
\label{tab:mechanism-ablation}
\end{table}

\section*{Limitations}

\paragraph{Dependence on issue quality.}
The evidence specification $C_q$ is compiled from the issue
description and repository structure, so its completeness is bounded
by how well the issue characterizes the underlying problem. When the
issue is vague or under-specified, $C_q$ may omit conditions that
matter for a correct fix or include irrelevant ones, weakening the
gate's signal. The issues in SWE-bench Verified are human-validated
and reasonably well-specified, which bounds this effect in our
experiments. On noisier real-world trackers, a preceding
issue-clarification step could improve $C_q$ quality.

\paragraph{Shared-model evidence operations.}
The evidence specification, gap assessment, and guidance are all
produced by the same model that drives the agent. A blind spot the
model has at action-selection time it may also have when assessing
evidence sufficiency. We partially mitigate this by grounding $C_q$
against repository structure rather than relying on the model's
parametric knowledge alone, but using a stronger or independent model
for evidence operations is a natural extension that \tool's design
already accommodates.

\paragraph{Evaluation scope.}
We evaluate on SWE-bench Verified with two models and two scaffolds.
While this establishes cross-model and cross-scaffold generality, it
does not cover other languages, task types such as feature addition,
or benchmarks beyond SWE-bench.
\section{Conclusion}

We presented \tool, an evidence-conditioned execution layer that
addresses premature commitment in coding agents by tracking
task-specific evidence and gating commitment actions whose supporting
conditions remain unsatisfied. On SWE-bench Verified, \tool improves
Pass@1 by 4.8--11.8 percentage points across two models and two
scaffolds while reducing token usage by up to 12.1\%, without
modifying the underlying agent. The results demonstrate that
controlling \emph{when} an agent may commit is an effective and
lightweight complement to improving \emph{what} it chooses to do.

\bibliography{aaai2027}


\end{document}